\documentclass{elsart5}

\usepackage{graphicx}
\usepackage{amsmath,amssymb}
\usepackage{dcolumn}
\usepackage{multirow}
\usepackage{bm}
\usepackage{epstopdf}
\usepackage{threeparttable}

\usepackage[numbers,square,sort&compress]{natbib}
\usepackage[hang]{subfigure}
\usepackage[colorlinks=true,citecolor=blue,linkcolor=blue]{hyperref}%

\newcommand{\dg}{$^\circ$ }

\begin{document}

\begin{frontmatter}

\title{Thermodynamic modeling of the Hf-Si-O system}

\author{Dongwon Shin\corauthref{cor1}}
\ead{dus136@psu.edu}
\corauth[cor1]{Corresponding author.}
\author{Raymundo Arr\'{o}yave}
\author{Zi-Kui Liu}%
\address{Department of Materials Science and Engineering,\\
The Pennsylvania State University, University Park, PA 16802, USA}

\begin{abstract}
The Hf-O system has been modeled by combining existing experimental data and
first-principles calculations results through the CALPHAD approach. Special
quasirandom structures of $\alpha$ and $\beta$ hafnium were generated to
calculate the mixing behavior of oxygen and vacancies. For the total energy of
oxygen, vibrational, rotational and translational degrees of freedom were
considered. The Hf-O system was combined with previously modeled Hf-Si and Si-O
systems, and the ternary compound in the Hf-Si-O system, HfSiO$_4$ has been
introduced to calculate the stability diagrams pertinent to the thin film
processing.
\end{abstract}

\end{frontmatter}

\section{Introduction}

The Hf-O system has been considered as one of the most important systems in
various industrial fields, such as nuclear materials and high
temperature/pressure materials. Hafnium dioxide is known as the least volatile
of all the oxides and its high melting point, extreme chemical inertness, and
high thermal neutron capture cross section make it suitable for use as control
rods or neutron shielding. All of these reasons can make HfO$_2$ a promising
refractory material for future nuclear applications\cite{1981Kom}.

Recently, the Hf-O system has even attracted considerable attention for
semi-conductor materials. The current gate oxide material (SiO$_2$ in general)
thickness in advanced complementary metal-oxide semiconductor (CMOS) integrated
circuits has continuously decreased and has reached the current process
limit\cite{2003Say}. One solution to further improve their performance is to
use alternative materials with higher dielectric constants (\emph{k}), such as
ZrO$_2$ and HfO$_2$\cite{1996Hub,2003Ram}. In this regard, thermodynamic
stability calculations results showed that the interface between HfO$_2$ and Si
is found to be stable with respect to the formation of silicides whereas the
ZrO$_2$/Si interface is not\cite{2002Gut}. The stability of the HfO$_2$/Si
interface make the use of this oxide even more promising.

The other important group IVA transition metals, such as Ti and Zr, which show
very similar behavior as Hf, with oxygen are modeled recently by
\citet{1999Wal} and \citet{2005Wan} respectively. All these systems commonly
have a wide oxygen solubility ranges in the hcp phase, up to 33 at.\%(Ti-O), 29
at.\%(Zr-O), and 20 at.\%(Hf-O) at room temperature, as derived from higher
temperature measurements.

In the present work, the Hf-O system has been modeled with the existing
experimental data and first-principles calculations results. Afterwards,
combining with the thermodynamic parameters of the Hf-Si and the Si-O binary
systems, the thermodynamic description of the Hf-Si-O ternary system is
obtained, and the stability diagrams pertinent to thin film processing, such as
the HfO$_2$-SiO$_2$ pseudo-binary, the isopleth of HfO$_2$-Si, and isothermal
sections are calculated.

\section{Experimental data}

\subsection{Phase diagram data}

\subsubsection{Hf-O}

Many investigations have been conducted to clarify the phase diagram of the
Hf-O system\cite{1963Rud,1965Dom,1976Rud,1973Ruh}. The main issues regarding
the phase diagram of the Hf-O system include: the extent of the $\alpha$-Hf
solid solution, the congruent melting of HfO$_2$, and the allotropic
transformations of HfO$_2$.

The phase diagrams suggested by \citet{1963Rud} and \citet{1965Dom} are quite
similar to each other, except the formation of the $\beta$-Hf phase and the
eutectic reaction around 37 at.\% oxygen of liquid. For the hafnium rich-side,
\citet{1963Rud} proposed a eutectic reaction at 2273K, while \citet{1965Dom}
suggested a peritectic reaction around 2523K. Similar to other group IVA
transition metals, such as Ti, and Zr, it is strongly believed that Hf also has
a peritectic reaction at the Hf-rich side. Another minor disagreement between
these two phase diagram determinations is the eutectic reaction, Liquid
$\rightarrow$ $\alpha$ + HfO$_2$. Rudy and Stecher suggested composition of
oxygen in the liquid as 40 at.\% O and 2453$\pm$40K, while Domgala and Ruh
proposed 37 at.\% O and $\sim$2473K.

For the $\alpha$-Hf solid solution, these two works\cite{1963Rud,1965Dom} are
in quite good agreement with each other. Rudy and Stecher found that
$\alpha$-Hf dissolves up to 20.5 at.\% oxygen at 1623K and that the solubility
range is almost independent of temperature; this shows consistency with
observations by Domagala and Ruh and they quoted a solubility of oxygen in
$\alpha$-Hf of 18.6 at.\% at 1273K.

\citet{1973Ruh} proposed a tentative phase diagram for the $\rm HfO_2$-rich
portion of the Hf-HfO$_2$ system on the basis of metallographic data. They
suggested the existence of solid solution regions for both cubic and tetragonal
phases deviated from the stoichiometric composition of HfO$_2$. Since the
important phases in the Hf-O system are the polymorphs of the $\rm HfO_2$
phases, i.e. monoclinic, tetragonal, and cubic, they have been studied
extensively\cite{1953Gel,1954Cur,1959Ada,1965Bog,1975Sta}. Allotropic
transformations of the HfO$_2$ phase have been well summarized by
\citet{1992Wan}.

\begin{figure}[htb]
\begin{center}
    \includegraphics[width=3.5in]{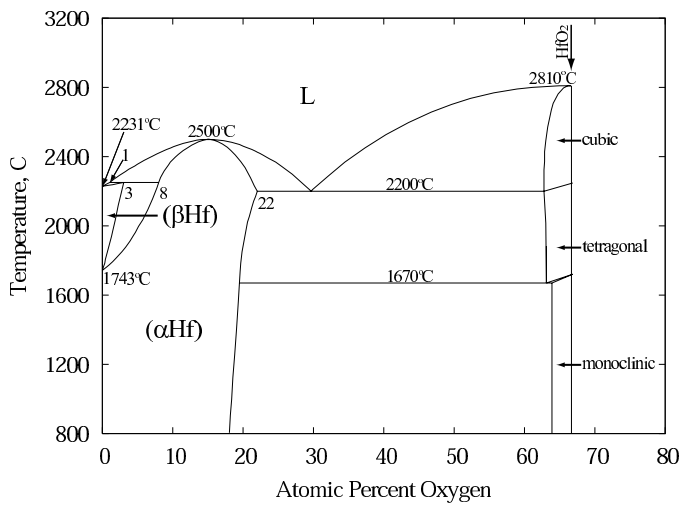}
\end{center}
\caption{%
Proposed phase diagram of the Hf-O system from \citet{1990Mas}.
}%
\label{fig:hfom}%
\end{figure}

The suggested phase diagram of the Hf-O system by \citet{1990Mas} is shown in
Fig. \ref{fig:hfom} .

\subsubsection{Hf-Si-O}

Not many studies have been conducted regarding the phase stabilities of the
Hf-Si-O ternary system. \citet{1982Spe} reported a ternary compound, Hafnon,
with the chemical formula of HfSiO$_4$ and the crystal structure as $I4_1/amd$.
One of the most important phase diagrams of the Hf-Si-O system is the
HfO$_2$-SiO$_2$ pseudo-binary that includes HfSiO$_4$ since the phase
stabilities of this pseudo-binary are pertinent to the processing of the
dielectric thin film. \citet{1969Par} determined the melting of HfSiO$_4$ at
2023$\pm$15K.

\subsection{Thermochemical data}

As discussed in the introduction, the Hf-O system has a wide range of oxygen
solubility in the hcp phase. Hirabayashi \emph{et al.} studied order/disorder
transformation of interstitial oxygen in hafnium\cite{1973Hir} around 10 $\sim$
20at.\% by electron microscopy and neutron and X-ray diffractions. This work
revealed that two types of interstitial superstructures are formed in the hypo-
and hyper-stoichiometric compositions near $\rm HfO_{1/6}$ below 700K which
have $R\bar{3}$ and $P\bar{3}1c$ symmetries, respectively.

For the completely disordered hcp phase at high temperature, \citet{1984Bou}
measured the partial molar enthalpy of solution of oxygen in $\alpha$-Hf solid
solution at 1323K as a function of oxygen content using a Tian-Calvet-type
microcalorimeter. Previously, it was almost impossible to measure the extremely
low oxygen pressure in equilibrium with hafnium-oxygen solutions. Therefore
derivation from the second law of thermodynamics was the only way to acquire
thermodynamic information of solid solution phases\cite{1963Kom,1963Sil}. The
major difficulty of direct measurement is making sure that all the hafnium
surface is accessible at the same time to oxygen. Boureau and Gerdanian
improved the accuracy of measurement by solving the geometrical effect of
specimen and oxygen contact. The observed phase boundary of $\alpha$-Hf in this
work is consistent with that of previous phase diagram
studies\cite{1963Rud,1965Dom} as O/Hf=0.255.

\section{First-principles calculations}

\subsection{Methodology}

The Vienna \emph{Ab initio} Simulation Package (VASP)\cite{1996Kre} was used to
perform the Density Functional Theory (DFT) electronic structure calculations.
The projector augmented wave (PAW) method\cite{1999Kre} was chosen and the
general gradient approximation (GGA)\cite{1992Per} was used to take into
account exchange and correlation contributions to the hamiltonian of the
ion-electron system. An energy cutoff of 500 eV was used to calculate the
electronic structures of all the compounds. 5,000 {\em k}-points per reciprocal
atom based on the Monkhorst-Pack scheme for the Brillouin-zone sampling was
used. The $k$-point meshes were centered at the $\Gamma$ point for the hcp
calculations.

\subsection{Ordered phases}

The ordered structures of the Hf-Si-O system calculated in this work can be
categorized into three groups. First, pure elements, i.e. hcp hafnium and
diamond silicon are calculated for the reference states. Second, hypothetical
compounds, the end-members of the $\alpha$, $\beta$ solid solutions
(HfO$_{0.5}$ and HfO$_3$), are also calculated. The stable compounds,
monoclinic HfO$_2$, quartz SiO$_2$, and the ternary compounds, HfSiO$_4$ are
calculated as well. The calculated results of the ordered structures are listed
in Table \ref{tbl:abinitio}. The enthalpy of those compounds are calculated
from Eqn. \ref{eqn:hfsio}:

\begin{align}
\label{eqn:hfsio} %
\Delta H^{Hf_xSi_yO_z}_{f} = & H(Hf_xSi_yO_z)-\tfrac{x}{x+y+z}H(Hf)\nonumber\\
                             & -\tfrac{y}{x+y+z}H(Si)-\tfrac{z}{x+y+z}H(O)
\end{align}

\noindent where $H$ corresponds to the enthalpies of the compound and reference
structures. The reference states for Hf and Si were the hcp and diamond
structures, respectively. In the case of condensed phases, the effects of
lattice vibrations and other degrees of freedom (i.e. electronic, magnetic) can
be neglected at low temperatures. Moreover, due to their rather small molar
volumes, the $PV$ contributions to their enthalpies ($H\equiv U+PV$) can also
be neglected. Thus their enthalpies $H$, can be replaced by the calculated
first-principles total energies at 0K in Eqn. \ref{eqn:hfsio}. For the oxygen
gas, the selected reference state was diatomic oxygen, O$_2$. In this case, the
contributions due to vibrational and translational degrees of freedom, as well
as the $PV$ work term, the molar volume of O$_2$ is much larger than that of
the condensed phases and cannot be neglected. In the following section, the
determination of the correct reference state for O$_2$ will be briefly
discussed.

\begin{table*}[htb]
\centering %
\caption{First-principles calculation results of pure elements,
hypothetical compounds ($\alpha$, $\beta$-Hf), and stable compounds (HfO$_2$,
SiO$_2$, and HfSiO$_4$). By definition, $\Delta H^f$ of pure elements are zero.
Reference states for all the compounds are SER.}
\label{tbl:abinitio}%
\begin{tabular}{lllllcc}
\hline %
Phases & Space & \multicolumn{3}{l}{Lattice parameters (\AA)} & Total energy & $\Delta H^f$\\
       & Group & a & b & c                                    & (eV/atom)    & (kJ/mol-atom)\\
\hline %
HCP\_A3 (Hf)              & $P6_{3}/mmc$ & 3.198 & 3.198 & 5.053 & -9.8320 & 0\\
Diamond\_A4 (Si)          & $Fd\bar{3}m$ & 5.468 & 5.468 & 5.468 & -5.4315 & 0\\
Gas (O$_2$)               &              &       &       &       & -4.7936 & 0\\
\hline %
$\alpha$-Hf (HfO$_{0.5}$) & $P\bar{3}m1$ & 3.225 & 3.225 & 5.150 & -9.9718 & -175.511 \\
$\beta$-Hf  (HfO$_3$)     & $Im\bar{3}m$ & 4.364 & 4.364 & 4.364 & -7.7253 & -161.308 \\
\hline %
Monoclinic(HfO$_2$)$\rm ^a$ %
                          & $P2_1/c$     & 5.135 & 5.194 & 5.314 & -10.2101 & -360.563\\
Quartz (SiO$_2$)          & $P3_221$     & 5.007 & 5.007 & 5.496 & -7.9581 & -284.809 \\
HfSiO$_4$                 & $I4_1/amd$   & 6.616 & 6.616 & 6.004 & -9.1024 & -324.453 \\
 &  &  &  &  &  & -1.769$\rm ^b$ \\
\hline %
\end{tabular}
\begin{tablenotes}
\item $\rm ^a$ $\beta$=99.56\dg
\item $\rm ^b$ Reference states are monoclinic (HfO$_2$) and quartz (SiO$_2$).
\end{tablenotes}
\end{table*}

\subsection{Oxygen calculation}

The selected reference state for oxygen corresponds to the diatomic molecule at
298K. Two oxygen atoms were placed in a 'big box' to model the oxygen molecule
gas and completely relaxed to find the lowest energy configuration. In order to
properly account for the net magnetic moment of this molecule, spin
polarization was considered. It was necessary to take into account the
contributions of vibrational, rotational and translational degrees of freedom
at finite temperature. Under the harmonic oscillator-rigid rotor approximation
at temperatures greater than the characteristic rotational temperature--- 2.07K
for O$_2$---the internal energy of the O$_2$ molecule is given by~\cite{McQ00}:

\begin{equation}
\label{eqn:E_diatomic} %
E(T) = k_BT\left(\frac{5}{2}+\frac{\Theta_{\nu}}{2}
+\frac{\Theta_{\nu}/T}{e^{\Theta_{\nu}/T}-1}\right)+E_{0}
\end{equation}

\noindent where $\Theta_{\nu}$ is the characteristic vibrational temperature.
The first term in Eqn. \ref{eqn:E_diatomic} corresponds to the contributions
due to translational and rotational degrees of freedom and the second and third
terms correspond to vibrational contributions. The last term, E$_0$,
corresponds to the energy of the ground electronic state at 0K. By including
the $PV=k_BT$ term of an ideal, non-interacting gas in Eqn.
\ref{eqn:E_diatomic}, the enthalpy of the diatomic O$_2$ molecule can be
obtained. The difference between the ground state electronic energy and the
enthalpy of O$_2$, per atom, is given by:

\begin{equation}
\label{eqn:H_diatomic} %
H(T)^{O_2}-E_{0}^{O_2}= \frac{k_BT}{2}\left(\frac{7}{2}+\frac{\Theta_{\nu}}{2}
+\frac{\Theta_{\nu}/T}{e^{\Theta_{\nu}/T}-1}\right)
\end{equation}

In the case of O$_2$, the characteristic vibrational temperature,
$\Theta_{\nu}$ is 2256K~\cite{McQ00}. At 298K, the value of
$H(T)^{O_2}-E_{0}^{O_2}$ is +0.0936 eV/atom or +9.03 kJ/mol-atom.

\subsection{Disordered phases: Special quasirandom structures}

In order to calculate the enthalpies of mixing of oxygen and vacancy for both
hcp and bcc phases in the Hf-O system, special quasirandom structures
(SQS)\cite{1990Zun} are used which are supercells with correlation functions as
close as possible to those of a completely random solution phase.

In order to introduce the special quasirandom structure, it is convenient to
understand the concept of \emph{correlation functions}. In any atomic
arrangements, the geometrical correlation between atoms in the structure can be
defined as to whether it is ordered or disordered. For a completely disordered
structure, the surrounding environment of an atom at any given sites should be
the same as all the other lattice sites. Correlation functions, $\prod_{k,m}$,
are defined as the products of site occupation numbers of different figures,
$k$, such as point, pair, triplet (when $k=1,2,3,\dots$) and so forth. The
correlation functions of each figure can be grouped together based on the
distance from a lattice site as m$^{th}$ nearest-neighbors.

For a binary system, a spin variable, $\sigma=\pm1$, can be assigned to
different types of atomic occupations and their products represent the
correlation functions of binary alloys. The correlation function of a random
alloy is simply described as $(2x-1)^k$ in the $A_{1-x}B_x$ substitutional
binary alloy, where $x$ is the composition. Once such a supercell satisfies the
correlation function of a target structure, it can be easily transferred to
other systems by simply switching the types of atoms in the structure.

The major drawback of SQS is that the concentration which can be calculated is
typically limited to 25, 50, and 75 at.\% since the correlation functions for
completely random structures other than those three compositions are almost
impossible to satisfy with a small number of atoms. In principle, one can find
a bigger supercell which has better correlation functions than smaller ones;
however, such a calculation requires expensive computing. On the other hand,
three data points from SQS calculations can clearly indicate the mixing
behavior of solution phases. Another disadvantage of SQS is that it cannot
consider the long range interaction since the size of the structure itself is
limited. It is reported that SQS works well with a system where short range
interactions are dominant\cite{2004Jia,2006Shi}.

Three different compositions, i.e. $y_O$ = 0.25, 0.5, and 0.75 with $y_O$
representing the mole fraction of oxygen in the hcp and bcc interstitial sites,
were considered and only two structures were generated in both phases since the
structures of $y_O=0.25$ and 0.75 are switchable to each other. For the
$\alpha$ solid solution,the  total number of lattice sites considered were 24,
36, and 48. Since only the mixing between oxygen and vacancies are considered
to generate a SQS for the hcp and bcc phases, the hafnium ions are excluded
from the correlation function calculations. Therefore, the total number of
oxygen and vacancies are 8, 12, and 16, respectively. For the $\beta$ solid
solution, the total number of lattice sites that were considered for the mixing
of oxygen and vacancy were 12 and 24 with total number of sites being 16 and
32, respectively. The complete descriptions of the SQS's for $\alpha$ and
$\beta$ solid solutions are listed in Tables \ref{tbl:hcpsqs_cor} and
\ref{tbl:bccsqs_cor}, and their correlation functions are given in Tables
\ref{tbl:alphaCF} and \ref{tbl:betaCF}. Finding bigger cells than these were
prohibited by the limited computing resources.

\begin{table*}[htb]
\centering %
\caption{Structural descriptions of the SQS-$N$ structures for the $\alpha$
solid solution. Lattice vectors and atom/vacancy positions are given in
fractional coordinates of the supercell. Atomic positions are given for the
ideal, unrelaxed hcp sites. Translated Hf positions are not listed. Original Hf
positions in the primitive cell are (0 0 0) and ($\tfrac{2}{3}$ $\tfrac{1}{3}$
$\tfrac{1}{2}$).}
\label{tbl:hcpsqs_cor}%
\begin{tabular}{ccccc}
\hline %
Oxygen \% & \multicolumn{3}{c}{50} & 75 \\
SQS-$N$   & 24 & 36 & 48           & 48 \\
\hline %
Lattice vectors %
 & $\left(\begin{array}{rrr} 
   1 &  0 & -1 \\
  -1 & -2 &  0 \\
  -3 &  0 & -1 \\
\end{array}\right)$
 & $\left(\begin{array}{rrr} 
   1 & -1 &  1 \\
  -2 & -2 &  1 \\
  -1 & -3 & -1 \\
\end{array}\right)$
 & $\left(\begin{array}{rrr} 
   0 & -2 &  0 \\
   0 &  0 & -2 \\
   4 &  2 &  0 \\
\end{array}\right)$
 & $\left(\begin{array}{rrr} 
   2 &  1 & -1 \\
   0 & -2 & -1 \\
  -2 &  1 & -2 \\
\end{array}\right)$ \\

O %
& $\begin{array}[t]{rrr}
  \tfrac{1}{3} &   \tfrac{1}{3} &  -\tfrac{3}{4} \\%
-1\tfrac{2}{3} & -1\tfrac{1}{3} &  -\tfrac{3}{4} \\%
 -\tfrac{2}{3} & -1\tfrac{1}{3} &  -\tfrac{3}{4} \\%
-2\tfrac{2}{3} & -1\tfrac{1}{3} & -1\tfrac{3}{4} \\%
\end{array}$ %
& $\begin{array}[t]{rrr} 
 -\tfrac{2}{3} & -2\tfrac{1}{3} &  \tfrac{1}{4} \\%
-1\tfrac{2}{3} & -2\tfrac{1}{3} & 1\tfrac{1}{4} \\%
-2\tfrac{2}{3} & -4\tfrac{1}{3} &  \tfrac{1}{4} \\%
 -\tfrac{2}{3} & -2\tfrac{1}{3} & 1\tfrac{1}{4} \\%
-1\tfrac{2}{3} & -4\tfrac{1}{3} &  \tfrac{1}{4} \\%
 -\tfrac{2}{3} & -4\tfrac{1}{3} &  \tfrac{1}{4} \\%
\end{array}$ %
& $\begin{array}[t]{rrr} 
  \tfrac{1}{3} & -1\tfrac{1}{3} &  -\tfrac{3}{4} \\%
 2\tfrac{1}{3} &  -\tfrac{1}{3} &  -\tfrac{3}{4} \\%
 3\tfrac{1}{3} &  -\tfrac{1}{3} &  -\tfrac{3}{4} \\%
 1\tfrac{1}{3} & -1\tfrac{1}{3} & -1\tfrac{3}{4} \\%
 2\tfrac{1}{3} &   \tfrac{2}{3} & -1\tfrac{3}{4} \\%
 3\tfrac{1}{3} &   \tfrac{2}{3} & -1\tfrac{3}{4} \\%
 2\tfrac{1}{3} &  -\tfrac{1}{3} & -1\tfrac{3}{4} \\%
 3\tfrac{1}{3} &  -\tfrac{1}{3} & -1\tfrac{3}{4} \\%
\end{array}$ %
& $\begin{array}[t]{rrr} 
 -\tfrac{2}{3} &   \tfrac{2}{3} & -2\tfrac{3}{4} \\%
  \tfrac{1}{3} &   \tfrac{2}{3} & -2\tfrac{3}{4} \\%
  \tfrac{1}{3} &  1\tfrac{2}{3} & -2\tfrac{3}{4} \\%
 -\tfrac{2}{3} &  -\tfrac{1}{3} & -1\tfrac{3}{4} \\%
 1\tfrac{1}{3} &  -\tfrac{1}{3} & -1\tfrac{3}{4} \\%
 -\tfrac{2}{3} &   \tfrac{2}{3} & -1\tfrac{3}{4} \\%
  \tfrac{1}{3} &   \tfrac{2}{3} & -1\tfrac{3}{4} \\%
 1\tfrac{1}{3} &   \tfrac{2}{3} & -1\tfrac{3}{4} \\%
 -\tfrac{2}{3} &  -\tfrac{1}{3} & -2\tfrac{3}{4} \\%
  \tfrac{1}{3} &  -\tfrac{1}{3} & -2\tfrac{3}{4} \\%
 -\tfrac{2}{3} & -1\tfrac{1}{3} & -1\tfrac{3}{4} \\%
  \tfrac{1}{3} &  -\tfrac{1}{3} &  -\tfrac{3}{4} \\%
\end{array}$ \\

\hline%

Va %
& $\begin{array}[t]{rrr} 
-1\tfrac{2}{3} &  -\tfrac{1}{3} & -1\tfrac{3}{4} \\%
-1\tfrac{2}{3} &  -\tfrac{1}{3} &  -\tfrac{3}{4} \\%
 -\tfrac{2}{3} &  -\tfrac{1}{3} &  -\tfrac{3}{4} \\%
-2\tfrac{2}{3} & -1\tfrac{1}{3} &  -\tfrac{3}{4} \\%
\end{array}$
& $\begin{array}[t]{rrr} 
-1\tfrac{2}{3} & -3\tfrac{1}{3} &   \tfrac{1}{4} \\%
 -\tfrac{2}{3} & -3\tfrac{1}{3} &   \tfrac{1}{4} \\%
-1\tfrac{2}{3} & -4\tfrac{1}{3} &  1\tfrac{1}{4} \\%
 -\tfrac{2}{3} & -1\tfrac{1}{3} &   \tfrac{1}{4} \\%
-1\tfrac{2}{3} & -3\tfrac{1}{3} &  1\tfrac{1}{4} \\%
 -\tfrac{2}{3} & -3\tfrac{1}{3} &  1\tfrac{1}{4} \\%
\end{array}$
& $\begin{array}[t]{rrr} 
  \tfrac{1}{3} &  -\tfrac{1}{3} &  -\tfrac{3}{4} \\%
 1\tfrac{1}{3} &  -\tfrac{1}{3} &  -\tfrac{3}{4} \\%
 1\tfrac{1}{3} & -1\tfrac{1}{3} &  -\tfrac{3}{4} \\%
 2\tfrac{1}{3} &   \tfrac{2}{3} &  -\tfrac{3}{4} \\%
 3\tfrac{1}{3} &   \tfrac{2}{3} &  -\tfrac{3}{4} \\%
  \tfrac{1}{3} &  -\tfrac{1}{3} & -1\tfrac{3}{4} \\%
 1\tfrac{1}{3} &  -\tfrac{1}{3} & -1\tfrac{3}{4} \\%
  \tfrac{1}{3} & -1\tfrac{1}{3} & -1\tfrac{3}{4} \\%
\end{array}$
& $\begin{array}[t]{rrr} 
  \tfrac{1}{3} &  -\tfrac{1}{3} & -1\tfrac{3}{4} \\%
  \tfrac{1}{3} & -1\tfrac{1}{3} & -1\tfrac{3}{4} \\%
 1\tfrac{1}{3} &   \tfrac{2}{3} &  -\tfrac{3}{4} \\%
-1\tfrac{2}{3} &  -\tfrac{1}{3} & -2\tfrac{3}{4} \\%
\end{array}$ \\
\hline %
\end{tabular}
\end{table*}

\begin{table*}[htb]
\centering %
\fontsize{11}{11pt}\selectfont %
\caption{Structural descriptions of the SQS-$N$ structures for the $\beta$
solid solution. Lattice vectors and atom/vacancy positions are given in
fractional coordinates of the supercell. Atomic positions are given for the
ideal, unrelaxed bcc sites. Translated Hf positions are not listed. The
original Hf position in the primitive cell is (0 0 0).}
\label{tbl:bccsqs_cor}%
\begin{tabular}{cccc}
\hline %
Oxygen \% & \multicolumn{2}{c}{50} & 75 \\
SQS-$N$   & 16 & 32                & 32 \\
\hline %
Lattice vectors %
 & $\left(\begin{array}{rrr} 
 0.5 &  0.5 &  1.5 \\
-0.5 &  1.5 &  0.5 \\
-0.5 & -1.5 & -0.5\\
\end{array}\right)$
 & $\left(\begin{array}{rrr} 
-1 &  0 &  0 \\
 0 &  1 & -1 \\
 0 & -2 & -2 \\
\end{array}\right)$
 & $\left(\begin{array}{rrr} 
 0 & 0 & 2 \\
 0 & 2 & 0 \\
-1 & 0 & 0\\
\end{array}\right)$ \\

O 
& $\begin{array}[t]{rrr}
 0   &  0   &  0.5 \\%
-0.5 &  1   &  1   \\%
-0.5 &  0   &  1   \\%
 0   & -0.5 &  1   \\%
 0   &  1   &  1.5 \\%
 0   &  0.5 &  1.5 \\%
\end{array}$ %
& $\begin{array}[t]{rrr} 
-1   &  0   & -0.5 \\%
-1   &  0.5 & -1.5 \\%
-0.5 &  0   & -1.5 \\%
-1   &  0.5 & -1   \\%
-0.5 &  0.5 & -1   \\%
-1   & -1   & -1.5 \\%
-1   & -0.5 & -2.5 \\%
-1   & -0.5 & -1   \\%
-0.5 &  0   & -2   \\%
-1   & -0.5 & -2   \\%
-0.5 & -0.5 & -2   \\%
-0.5 & -1   & -2   \\%

\end{array}$ %
& $\begin{array}[t]{rrr} 
-0.5 & 2   & 1.5 \\%
-0.5 & 1   & 1.5 \\%
-1   & 1   & 1.5 \\%
-1   & 0.5 & 1.5 \\%
 0.5 & 2   & 0.5 \\%
 1   & 2   & 0.5 \\%
-1   & 1.5 & 0.5 \\%
-1   & 1   & 0.5 \\%
-1   & 0.5 & 0.5 \\%
-1   & 1.5 & 1 \\%
-0.5 & 1.5 & 1 \\%
-0.5 & 1   & 1 \\%
-1   & 0.5 & 1 \\%
-0.5 & 0.5 & 1 \\%
-0.5 & 2   & 1 \\%
 0.5 & 1.5 & 2 \\%
-0.5 & 1   & 2 \\%
-1   & 0.5 & 2 \\%
\end{array}$ \\

\hline%

Va 
& $\begin{array}[t]{rrr} 
-0.5 &  1   & 0.5 \\%
-0.5 & -0.5 & 0   \\%
-0.5 & -1   & 0   \\%
-0.5 &  0   & 0.5 \\%
 0   &  0.5 & 1   \\%
-0.5 &  0.5 & 1   \\%
\end{array}$
& $\begin{array}[t]{rrr} 
-0.5 & -1   & -2.5 \\%
-1   & -1   & -2.5 \\%
-1   & -1.5 & -2.5 \\%
-0.5 &  0   & -0.5 \\%
-1   & -1.5 & -2   \\%
-0.5 & -1.5 & -2   \\%
-0.5 & -1   & -3   \\%
-1   &  0   & -1.5 \\%
-1   & -0.5 & -1.5 \\%
-0.5 &  0   & -1   \\%
-0.5 & -1   & -1.5 \\%
-0.5 & -0.5 & -1   \\ %
\end{array}$
& $\begin{array}[t]{rrr} 
-1   & 2   & 1.5 \\%
-1   & 1.5 & 1.5 \\%
-0.5 & 1   & 0.5 \\%
-1   & 1.5 & 2   \\%
-0.5 & 0.5 & 2   \\%
-0.5 & 2   & 2   \\%
\end{array}$ \\
\hline %
\end{tabular}
\end{table*}

\begin{table}[htb]
\centering %
\caption{\label{tbl:alphaCF}Pair and multi-site correlation functions of
SQS-$N$ structures for $\alpha$ solid solution when the c/a ratio is ideal. The
number in the square bracket next to $\overline{\Pi}_{k,m}$ is the number of
equivalent figures at the same distance in the structure.}
\begin{tabular}{lcccccc}
\hline %
Oxygen \%  & \multicolumn{4}{c}{50} & \multicolumn{2}{c}{75}\\
SQS-$N$    & Random & 24 & 36 & 48  & Random & 48\\
\hline
$\overline{\Pi}_{2,1}$[3] & 0 & 0        & 0        &  0       & 0.25  & 0.25 \\
$\overline{\Pi}_{2,2}$[1] & 0 & 0        & 0        &  0       & 0.25  & 0.25 \\
$\overline{\Pi}_{2,3}$[3] & 0 & 0        & 0        &  0       & 0.25  & 0.25 \\
$\overline{\Pi}_{2,4}$[6] & 0 & 0        & -0.16667 &  0       & 0.25  & 0.20833\\
$\overline{\Pi}_{2,5}$[3] & 0 & 0        & -0.11111 &  0       & 0.25  & 0.25 \\
$\overline{\Pi}_{2,6}$[3] & 0 & -0.16667 &  0.11111 &  0       & 0.25  & 0.41667\\
$\overline{\Pi}_{3,1}$[2] & 0 & 0        & -0.33333 & -0.25    & 0.125 & 0.25\\
$\overline{\Pi}_{3,1}$[6] & 0 & 0        &  0.11111 & -0.08333 & 0.125 & 0.08333\\
$\overline{\Pi}_{3,2}$[2] & 0 & 0        &  0.33333 &  0.25    & 0.125 & 0.25\\
\hline %
\end{tabular}
\end{table}

\begin{table}[htb]
\centering %
\caption{\label{tbl:betaCF}Pair and multi-site correlation functions of SQS-$N$
structures for $\beta$ solid solution. The number in the square bracket next to
$\overline{\Pi}_{k,m}$ is the number of equivalent figures at the same distance
in the structure.}
\begin{tabular}{lccccc}
\hline %
Oxygen \%  & \multicolumn{3}{c}{50} & \multicolumn{2}{c}{75}\\
SQS-$N$    & Random & 16 & 32       & Random & 32\\
\hline
$\overline{\Pi}_{2,1}$[6]  & 0 & 0        & 0       & 0.25  & 0.25 \\
$\overline{\Pi}_{2,2}$[12] & 0 & 0        & 0       & 0.25  & 0.25 \\
$\overline{\Pi}_{2,3}$[12] & 0 & 0        & 0       & 0.25  & 0.25 \\
$\overline{\Pi}_{2,4}$[6]  & 0 & 0.16667  & 0       & 0.25  & 0.33333\\
$\overline{\Pi}_{2,4}$[3]  & 0 & 0        & 0       & 0.25  & 0.33333 \\
$\overline{\Pi}_{2,5}$[24] & 0 & -0.29167 & 0       & 0.25  & 0.25\\
$\overline{\Pi}_{2,6}$[24] & 0 & -0.08333 & 0       & 0.25  & 0.25\\
$\overline{\Pi}_{3,1}$[12] & 0 & -0.16667 & 0       & 0.125 & 0.16667\\
$\overline{\Pi}_{3,2}$[8]  & 0 & 0        & 0       & 0.125 & 0\\
$\overline{\Pi}_{3,3}$[48] & 0 & 0.08333  & 0.08333 & 0.125 & 0.125\\
\hline %
\end{tabular}
\end{table}

The generated SQS's were either fully relaxed, or relaxed without allowing
local ion relaxations, i.e.  only volume for bcc and volume as well as c/a
ratio for hcp were optimized. Theoretically, all the first-principles
calculations should be fully relaxed to find the lowest energy configurations.
However, the structure should lie on the energy curve $vs.$ geometrical degree
of freedom of the same \emph{phase}. If the fully relaxed final structure does
not have the same crystal structure as initial input, it is not the phase of
interest any longer. Thus it is necessary to force the structure to keep its
parent symmetry.

The calculated results of $\alpha$ and $\beta$ solid solution phases are listed
in Table \ref{tbl:sqshcp} and \ref{tbl:sqsbcc}. More detailed discussion of
calculation results of special quasirandom structures and optimized results are
found in a later section.

\begin{table}[htb]
\caption{First-principles calculations results of $\alpha$-Hf special
quasirandom structures. $FR$ and $SP$ represent 'Fully Relaxed' and 'Symmetry
Preserved', respectively. Oxygen atoms are excluded for the symmetry check.}
\label{tbl:sqshcp}%
\begin{tabular}{cccccccc}
\hline %
Oxygen & \multicolumn{3}{c}{Atoms} & Symmetry & Space & %
Total Energy & $\Delta H^{mix}$    \\
\%     & Hf & O & Va               &          & Group & %
(eV/atom)    & (kJ/mol-atom) \\ %
\hline %
25 & 32 & 4  & 12 & FR & $P6_{3}/mmc$ & -9.9140 & -61.9280\\
   & 32 & 4  & 12 & SP & $P6_{3}/mmc$ & -9.9072 & -61.2715\\
50 & 16 & 4  & 4  & FR & $P6_{3}/mmc$ & -9.9590 & -109.481\\
   & 16 & 4  & 4  & SP & $P6_{3}/mmc$ & -9.9444 & -108.075\\
   & 24 & 6  & 6  & FR & $P6_{3}/mmc$ & -9.9560 & -109.187\\
   & 24 & 6  & 6  & SP & $P6_{3}/mmc$ & -9.9418 & -107.818\\
   & 32 & 8  & 8  & FR & $P6_{3}/mmc$ & -9.9564 & -109.230\\
   & 32 & 8  & 8  & SP & $P6_{3}/mmc$ & -9.9413 & -107.768\\
75 & 32 & 12 & 4  & FR & $P6_{3}/mmc$ & -9.9755 & -146.428\\
   & 32 & 12 & 4  & SP & $P6_{3}/mmc$ & -9.9541 & -144.356\\
\hline %
\end{tabular}
\end{table}

\begin{table}[htb]
\caption{First-principles calculations results of $\beta$-Hf special
quasirandom structures. $FR$ and $SP$ represent 'Fully Relaxed' and 'Symmetry
Preserved', respectively. Oxygen atoms are excluded for the symmetry check.}
\label{tbl:sqsbcc}%
\begin{tabular}{cccccccc}
\hline %
Oxygen & \multicolumn{3}{c}{Atoms} & Symmetry & Space & %
 Total Energy & $\Delta H^{mix}$ \\
\%     & Hf & O & Va               &          & Group & %
(eV/atom)   & (kJ/mol-atom) \\
\hline %
25 & 8 & 6  & 18 & FR & $Pm$         & -9.9280  & -227.299\\
   & 8 & 6  & 18 & SP & $Im\bar{3}m$ & -9.0165  & -139.345\\
50 & 4 & 6  & 6  & FR & $C2/m$       & -10.0088 & -315.520\\
   & 4 & 6  & 6  & SP & $Im\bar{3}m$ & -8.5718  & -176.866\\
   & 8 & 12 & 12 & FR & $Pm$         & -9.9636  & -311.160\\
   & 8 & 12 & 12 & SP & $Im\bar{3}m$ & -8.6081  & -180.370\\
75 & 8 & 18 & 6  & FR & $Pm$         & -9.4727  & -307.100\\
   & 8 & 18 & 6  & SP & $Im\bar{3}m$ & -8.1752  & -181.908\\
\hline %
\end{tabular}
\end{table}

\section{Thermodynamic modeling}

\subsection{Hf-O}

Seven phases are modeled in the Hf-O system:  hcp, bcc, ionic liquid, gas, and
three polymorphs of HfO$_2$: monoclinic, tetragonal, and cubic phases. Detailed
discussions of individual phases are given below.

\subsubsection{HCP and BCC}

It has been reported that the stable solid phases of group IVA transition
metals (Ti, Zr, and Hf), the hcp and bcc phases, dissolve oxygen interstitially
into their octahedral sites\cite{1997Tsu}. The solid solutions of Hf-O are
modeled with the two sublattice model, with one sublattice occupied only by
hafnium and the other one occupied by both oxygen and vacancies:

$$ (Hf)_1(O,Va)_c$$

\noindent where $c$ corresponds to the ratio of interstitial sites to normal
sites in each structure. For the hcp phase, the ratio is derived as
$c=\frac{1}{2}$ from the pure hcp sublattice model, consistent with the
previous thermodynamic modeling of Ti-O\cite{1999Wal} and Zr-O\cite{2002Arr}.
For the bcc phase, the stoichiometric ratio $c$ is equal to 3 as in the
conventional bcc phase.

The Gibbs energies of the hcp and bcc phase can be described as:

\begin{align}
G^{HCP,BCC}_{m} = & y_O{}^0G_{HfO_c}+y_{Va}{}^0G_{HfVa_c} \nonumber\\%
                  & +cRT(y_O\ln(y_O)+y_{Va}\ln(y_{Va}))+G^{ex}_m\\
G^{ex}_m        = & y_Oy_{Va} \sum^{k}_{k=0} {}^kL_{(Hf:O,Va)}(T)(y_O-y_{Va})^k
\end{align}

\noindent where ${}^oG_{HfO_c}$ is the standard Gibbs energy of the
hypothetical oxide $\rm HfO_c$, which is one of the end members that
establishes the reference surface for this model; ${}^0G_{HfVa_c}$ corresponds
to the standard Gibbs energy of the pure bcc and hcp phases and the chemical
interaction for oxygen and vacancies in the second sublattice is given by
$y_Oy_{Va} \sum {}^kL_{(Hf:O,Va)}(T)(y_O-y_{Va})^k$. This is identical to a
Redlich-Kister polynomial\cite{1948Red}.

\subsubsection{Ionic liquid}

The liquid phase region goes from pure liquid hafnium to stoichiometric liquid
HfO$_2$. The ionic two-sublattice model is used for the liquid
phase\cite{1985Hil}:

\begin{equation*}
(C^{+v_i}_{i})_P(A^{-v_j}_{j},Va,B^{0}_{k})_Q
\end{equation*}

\noindent where $C^{+v_i}_{i}$ corresponds to the cation, $i$, with valence,
$+v_i$; $A_j$ to the anion, $j$, with valence, $-v_j$; Va are hypothetical
vacancies added for electro-neutrality when the liquid is away from
stoichiometry, having a valence equal to the average charge of the cation, Q;
and $B^0_K$ represents any neutral component dissolved in the liquid. The
numbers of sites in the sublattices, $P$ and $Q$, are varied in such a way that
electro-neutrality for all compositions is ensured with $y$ being the site
fractions:

\begin{align}
P&=\sum_j v_jy_{A_j}+Qy_{Va}\\
Q&=\sum_i v_iy_{C_i}
\end{align}

For the particular case of the Hf-O system, this two-sublattice model can be
further simplified:

\begin{equation*}
(Hf^{+4})_{4-2y_{O^{-2}}}(O^{-2},Va^{-4})_4
\end{equation*}

The Gibbs energy expression for this system is:

\begin{align}
G^{\rm Ionic~liquid}_m =
            & y_{Hf^{+4}}y_{O^{-2}}{}^0G^L_{(Hf^{+4})_2(Va^{-4})_4} \nonumber\\%
            & +4y_{Hf^{+4}}y_{Va^{-4}}{}^0G_{(Hf^{+4})(Va^{-4})}\nonumber\\
            & + RT(4-2y_{O^{-2}})(y_{Hf^{+4}}\ln(y_{Hf{+4}}))\nonumber\\
            & + RT(4)(y_{O^{-2}}\ln(y_{O^{-2}})+y_{Va^{-4}}\ln(y_{Va^{-4}}))\nonumber\\
            & + G^{ex}_m\\
G^{ex}_m  = & y_{Hf^{+4}}y_{O^{-2}}y_{Va^{-4}} %
            \sum^k_{k=0}{}^kL_{(Hf^{+4}:O^{-2},Va^{-4})}(T)\nonumber\\
            &\times(y_{O^{-2}}-y_{Va^{-4}})^k
\end{align}

\noindent where ${}^0G^L_{(Hf^{+4})_2(Va^{-4})_4}$ corresponds to the standard
Gibbs energy for two moles of liquid $HfO_2$; ${}^0G_{(Hf^{+4})(Va^{-4})}$ is
the standard Gibbs energy for pure hafnium liquid and
${}^kL_{(Hf^{+4}:O^{-2}:Va^{-4})}$ corresponds to the excess chemical
interaction parameters between oxygen and vacancies in the second sublattice.

\subsubsection{Gas}

To describe the oxygen-rich side of the Hf-O system, the gas phase was included
in the calculation. The ideal gas model was used and the following six species
were considered:
$$
\rm (O,O_2,O_3,Hf,HfO,HfO_2)
$$

The Gibbs energy of the gas phase can be described as:
\begin{align}
\rm G^{Gas}_m =~%
          & y_O({}^0G_O+RT\ln(P))+y_{O_2}({}^0G_{O_2}+RT\ln(P))\nonumber\\
          & +y_{O_3}({}^0G_{O_3}+RT\ln(P))\nonumber\\
          & +y_{Hf}({}^0G_{Hf}+RT\ln(P))\nonumber\\
          & +y_{HfO}({}^0G_{HfO}+RT\ln(P))\nonumber\\
          & +y_{HfO_2}({}^0G_{HfO_2}+RT\ln(P))\nonumber\\
          & +RT(y_O\ln(y_O)+y_{O_2}\ln(y_{O_2})+y_{O_3}\ln(y_{O_3}))\nonumber\\
          & +RT(y_{Hf}\ln(y_{Hf})+y_{HfO}\ln(y_{HfO})\nonumber\\
          & +y_{HfO_2}\ln(y_{HfO_2}))
\end{align}

\noindent with $y$ being the mole fraction of species in the gas phase.

\subsubsection{Polymorphs of HfO$_2$}

Thermodynamic descriptions of three polymorphs of HfO$_2$ have been obtained
from the SSUB database\cite{1999SGTE}. For simplicity, all three phases are
modeled as line compounds and the transformation temperatures for monoclinic
$\rightarrow$ tetragonal $\rightarrow$ cubic $\rightarrow$ liquid are 2100,
2793, and 3073K, respectively.

\subsection{Si-O}

\begin{figure}[htb]
\begin{center}
    \includegraphics[width=3.35in]{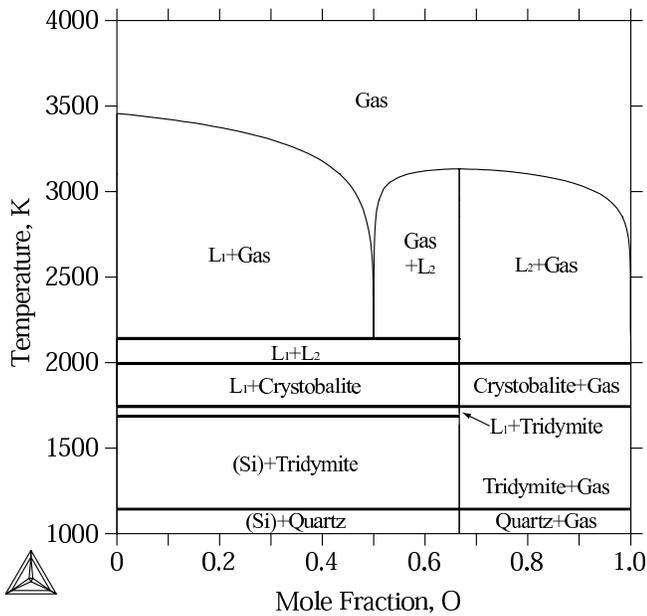}
\end{center}
\caption{%
Calculated Si-O phase diagram from Hallstedt\cite{1992Hal}.
}%
\label{fig:sio}
\end{figure}

The Si-O system has been modeled by \citet{1992Hal} with an ionic liquid model.
Three different polymorphs of silicon dioxides, quartz, tridymite, and
cristobalite are included in the system. The Si-O phase diagram is given in
Fig. \ref{fig:sio}.

\subsection{Hf-Si}

\begin{figure}[htb]
\begin{center}
    \includegraphics[width=3.35in]{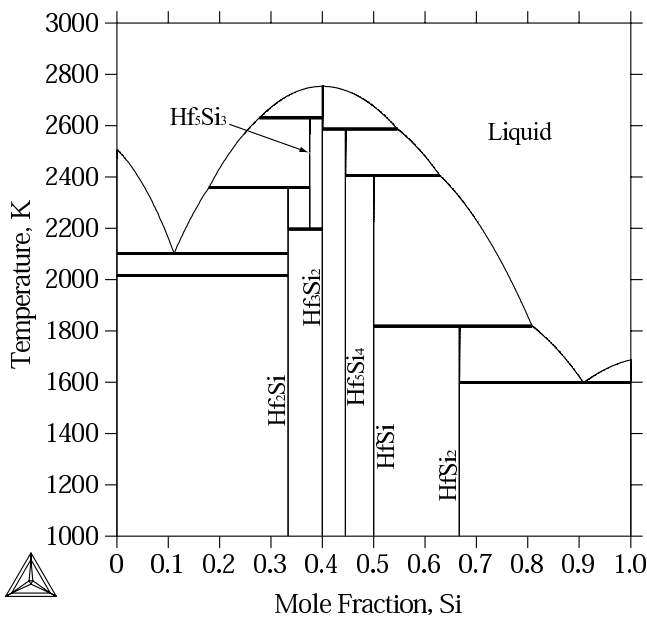}
\end{center}
\caption{%
Calculated Hf-Si phase diagram from Zhao \emph{et al}\cite{2000Zha}.
}%
\label{fig:hfsi}
\end{figure}

The Hf-Si system has been extensively studied and modeled by Zhao \emph{et
al}\cite{2000Zha}. Six intermetallic compounds, $\rm Hf_2Si$, $\rm Hf_5Si_3$,
$\rm Hf_3Si_2$, $\rm Hf_5Si_4$, $\rm HfSi$, and $\rm HfSi_2$ are present. The
calculated phase diagram of Hf-Si system is given in Fig. \ref{fig:hfsi}.

\subsection{Hf-Si-O}

In order to be combined with the Hf-O and Si-O systems, the liquid phase of the
Hf-Si system was converted to an ionic liquid in the present work. Hf$^{+4}$
and Si$^{+4}$ are in the first sublattice of the ionic liquid phase and
vacancies have been introduced into the second sublattice for the
electro-neutrality. The interaction parameters for the liquid phase from
\citet{2000Zha} were used for the mixing of Hf$^{+4}$ and Si$^{+4}$ in the
first ionic liquid sublattice.

The ternary compound HfSiO$_4$ has been introduced. Due to the lack of
experimental data, HfSiO$_4$ has been modeled as a stoichiometric compound.

\section{Results and discussion}

Based on the existing experimental data and the first-principles calculations
results, the model parameters for the Hf-O and Hf-Si-O systems are evaluated.
The PARROT module in the Thermo-Calc software has been used\cite{2002And}.

\begin{figure}[htb]
\begin{center}
    \includegraphics[width=3.35in]{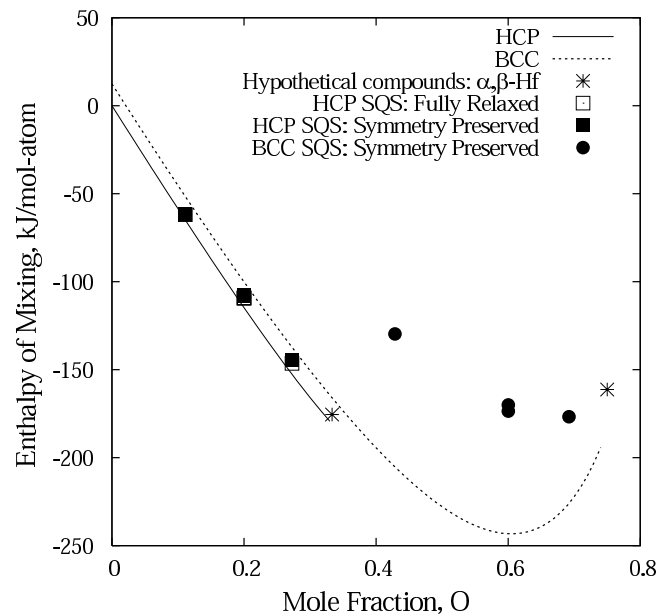}
\end{center}
\caption{%
First-principles calculations results of hypothetical compounds (HfO$_{0.5}$
and HfO$_3$) and special quasirandom structures for $\alpha$ and $\beta$ solid
solutions with the evaluated results. Reference states for Hf of $\alpha$ and
$\beta$ solid solutions are given as hcp. Fully relaxed calculations of $\beta$
solid solution have been excluded from this comparison since the calculation
results completely lost their bcc symmetry.
}%
\label{fig:sqs}%
\end{figure}

The enthalpy of mixing of the hcp and bcc phases calculated from the present
thermodynamic modeling are shown together with the result of first-principles
calculations in Fig. \ref{fig:sqs}. For the $\alpha$ solid solution, they agree
well with each other. As shown in Table \ref{tbl:sqshcp}, all first-principles
calculation results retained their original symmetry as hcp and the calculation
results are quite well converged with respect to the size of SQS. On the other
hand, results for the $\beta$ phase are not as good as those for the $\alpha$
phase. The difference between fully relaxed and symmetry preserved calculations
is 140kJ/mol-atom at most and all the fully relaxed calculations could not
maintain the bcc symmetry.

\begin{figure}[htb]
\begin{center}
    \includegraphics[width=3.35in]{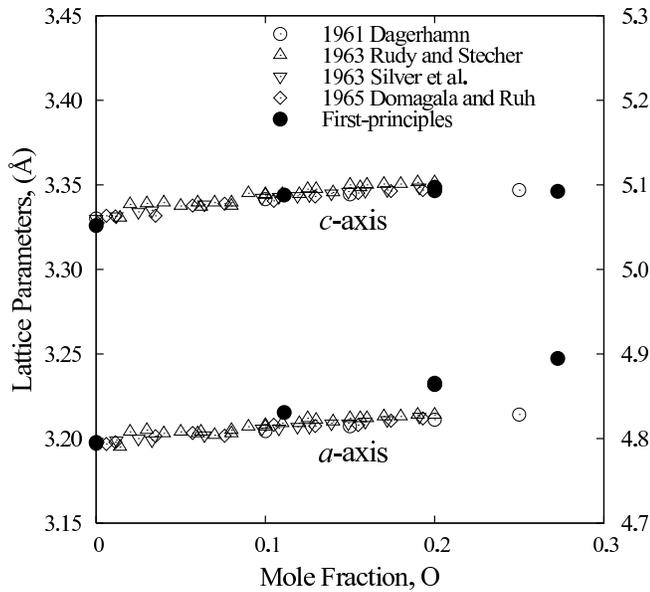}
\end{center}
\caption{%
Calculated lattice parameters of $\alpha$-Hf with experimental
data\cite{1963Rud,1965Dom,1963Sil,1961Dag}. Scale for $a$-axis is left and for
$c$ is right.
}%
\label{fig:lattice}%
\end{figure}

Such first-principles calculations results can be validated by comparing their
calculated crystallographic information with experimental measurements. In Fig.
\ref{fig:lattice}, the calculated lattice parameters of $\alpha$-Hf, both $a$
and $c$, are compared with experimental data and quite satisfactorily agree
with each other.

\begin{figure}[htb]
\begin{center}
    \includegraphics[width=3.35in]{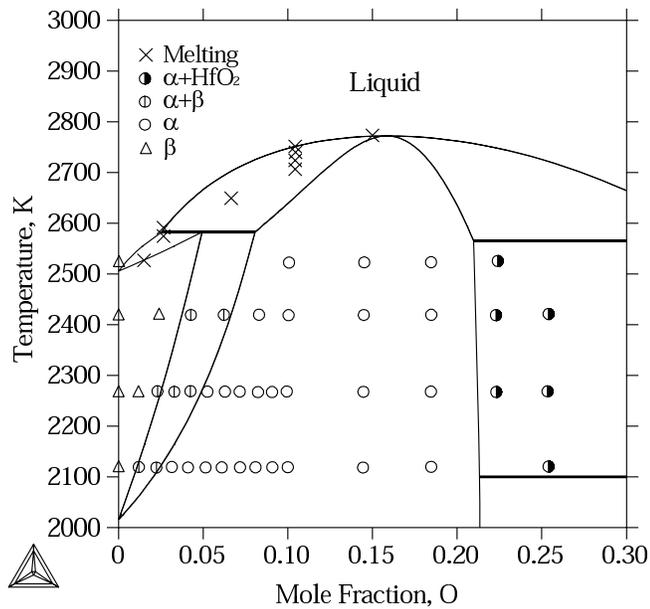}
\end{center}
\caption{%
Calculated Hf-rich side of the Hf-O phase diagram with experimental data from
\citet{1965Dom}.
}%
\label{fig:hfo2}%
\end{figure}

In Fig. \ref{fig:hfo2}, the Hf-rich side of the Hf-O phase diagram is shown.
The congruent melting of $\alpha$-Hf and the peritectic reaction are reproduced
correctly. The calculated phase region shows quite good agreement with the
X-ray phase identification results of \citet{1965Dom}.

\begin{figure}[htb]
\begin{center}
    \includegraphics[width=3.35in]{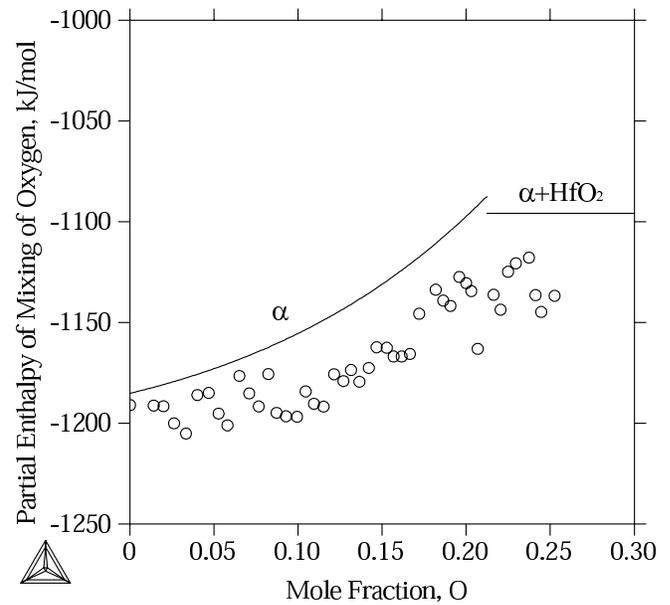}
\end{center}
\caption{%
Calculated partial enthalpy of mixing of oxygen in the $\alpha$-Hf with
experimental data\cite{1984Bou} at 1323K.
}%
\label{fig:hpo}%
\end{figure}

The calculated partial enthalpy of mixing of oxygen in the $\alpha$-Hf is shown
in Fig. \ref{fig:hpo} and compared with experimental data\cite{1984Bou}. As
mentioned before, the accuracy of measurement has been improved compared to the
previous experiments. However, it is still quite difficult to measure the low
oxygen pressure.

\begin{figure}[htb]
\begin{center}
    \includegraphics[width=3.35in]{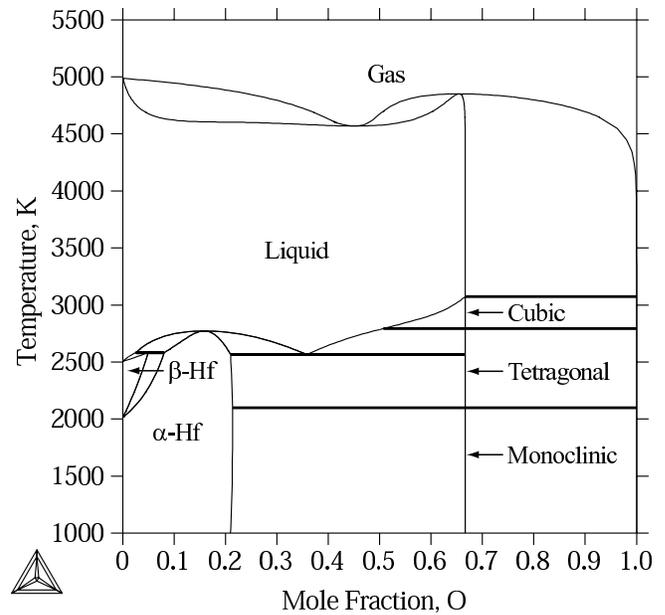}
\end{center}
\caption{%
Calculated Hf-O phase diagram.
}%
\label{fig:hfo}%
\end{figure}

The calculated phase diagram of the entire Hf-O system is shown in Fig.
\ref{fig:hfo} with the gas phase included.

\begin{figure}[htb]
\begin{center}
    \includegraphics[width=3.35in]{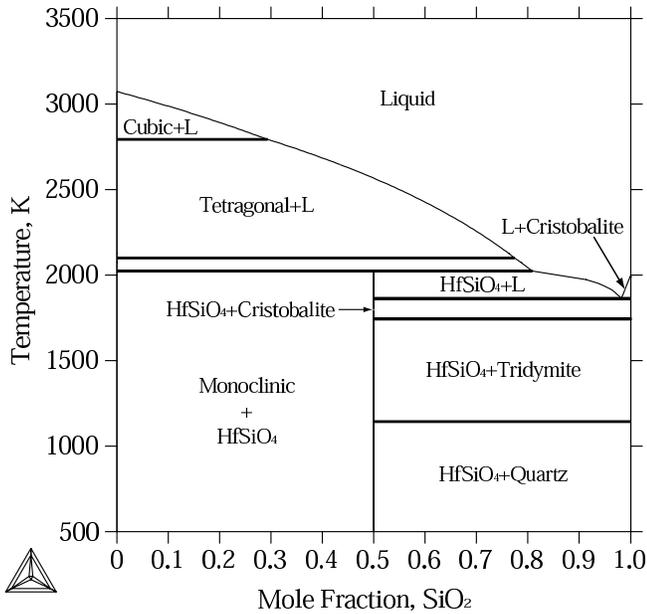}
\end{center}
\caption{%
Calculated HfO$_2$-SiO$_2$ pseudo-binary phase diagram.
}%
\label{fig:pseudo}%
\end{figure}

In the present work, the ternary liquid phase is extrapolated from the
binaries. The enthalpy of formation of HfSiO$_4$ is obtained from
first-principles calculations and the entropy of formation is evaluated from
its preitectic reaction, Liquid + HfO$_2$ (Monoclinic) $\rightarrow$ $\rm
HfSiO_4$, at 2023K as reported by Parfenenko \emph{et al}\cite{1969Par}. The
pseudo-binary phase diagram of HfO$_2$-SiO$_2$ is calculated and shown in Fig.
\ref{fig:pseudo}.

As discussed in the introduction, HfO$_2$ is a promising candidate to replace
SiO$_2$ as the gate dielectric in CMOS transistors due to its high dielectric
constant and compatibility with Si in comparison with ZrO$_2$\cite{2002Gut}. In
general, during the fabrication of such devices, the films are subjected to
temperatures around 1273K for a short period of time\cite{2003Ram}. Thus, it is
quite essential to understand the thermodynamic stability of
HfO$_2$/SiO$_2$/Si.

\begin{figure*}[htb]
\begin{center}
    \subfigure[500K]{
        \label{fig:isot500}%
        \includegraphics[width=3.35in]{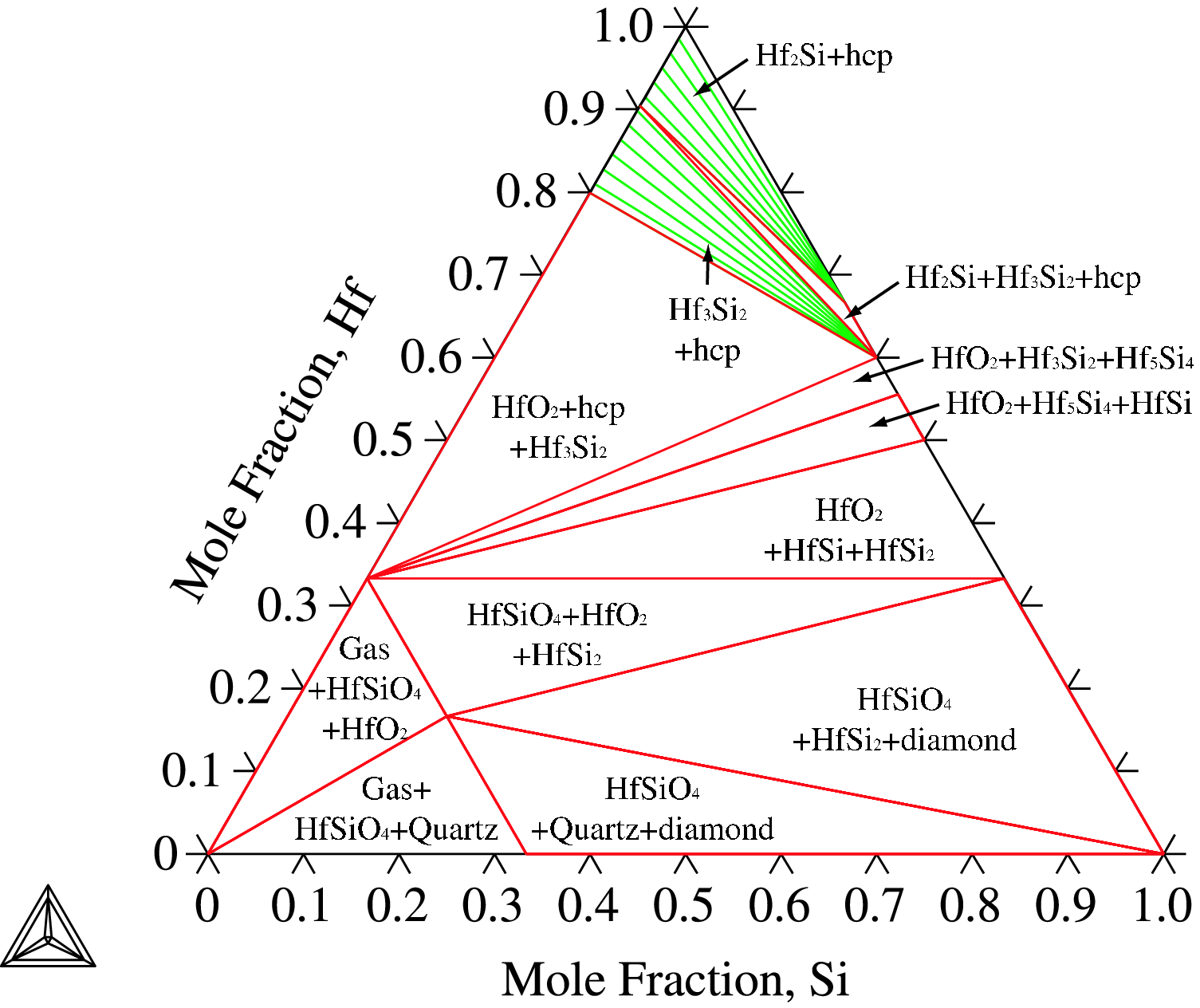}
    }
    \subfigure[1000K]{
        \label{fig:isot1000}%
        \includegraphics[width=3.35in]{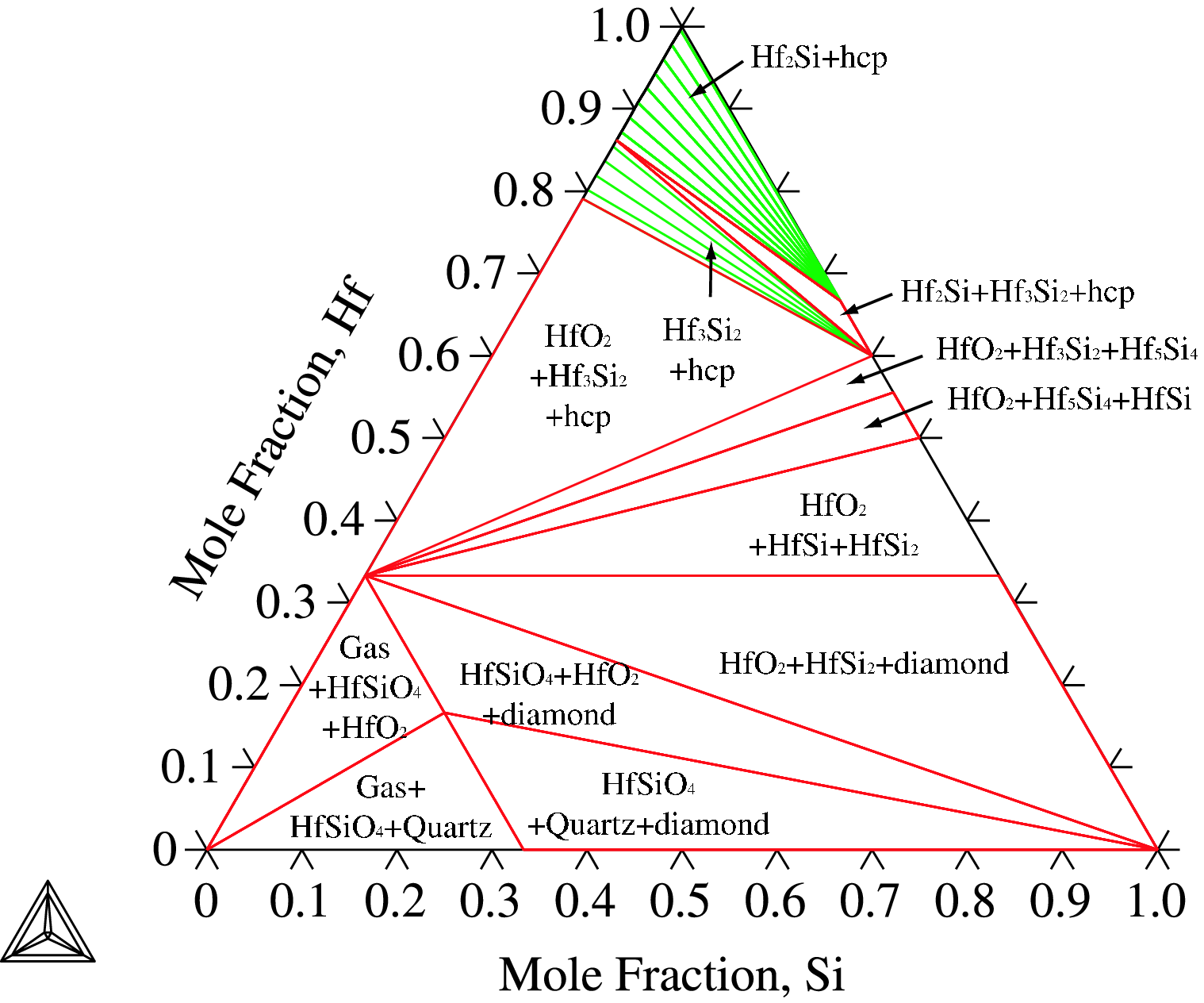}
    }
\end{center}
\caption{%
Calculated isothermal section of Hf-Si-O at (a) 500K and (b) 1000K at 1 atm.
Tie lines are drawn inside the two phase regions. The vertical cross section
between HfO$_2$ and Si is the isopleth in Figure \ref{fig:isop}.
}%
\label{fig:isot} %
\end{figure*}

From the evaluated thermodynamic database of the Hf-Si-O, the isothermal
sections of the Hf-Si-O system can be readily calculated to study the stability
of the HfO$_2$/Si interface. Two different temperatures are selected for the
calculations and they are 500K for low temperature processing, such as mist
deposition method and rapid thermal processing\cite{2004Cha} and 1000K, typical
temperature for the epitaxial growth of oxides deposition. Calculated
isothermal sections of the Hf-Si-O system at 500K and 1000K are shown in Fig.
\ref{fig:isot500} and \ref{fig:isot1000}, respectively. The two three-phase
regions, HfSiO$_4$+HfO$_2$+Hf$_2$Si and HfSiO$_4$+diamond+Hf$_2$Si, in the 500K
isothermal section should be noticed with respect to the stability of
HfO$_2$/Si interface. Since those regions are intersected by the line
connecting HfO$_2$ and Si, HfSi$_2$ can be found in the fabrication of
polySi/HfO$_2$ gate stack Metal Oxide Semiconductor Field Effect Transistor
(MOSFET) on bulk Si at 500K, while the 1000K calculation result clearly shows
that HfO$_2$ is stable with the Si substrate.

\begin{figure}[htb]
\begin{center}
    \includegraphics[width=3.35in]{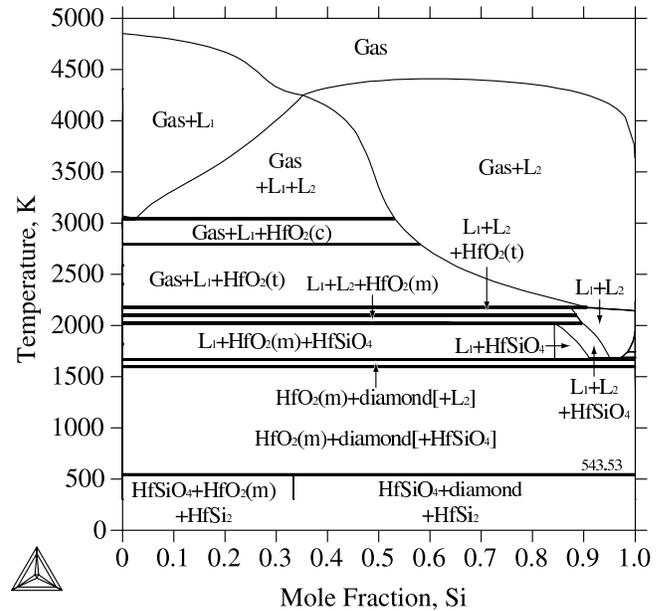}
\end{center}
\caption{%
Calculated isopleth of HfO$_2$-Si at 1 atm. Hafnium dioxide is left and silicon
is right. Polymorphs of HfO$_2$, monoclinic, tetragonal, and cubic, are given
in parentheses. The phases in the bracket are zero amount.
}%
\label{fig:isop}%
\end{figure}

A vertical cross section of isothermal sections, the isopleth between
HfO$_2$-Si is calculated in order to investigate the stability range of
HfSi$_2$ in the HfO$_2$/Si interface and is given in Fig. \ref{fig:isop}. It
should be noted that HfSiO$_4$ at the low temperature range is zero amount. The
calculated result shows that HfSi$_2$ becomes stable below 543.53K. This result
is in agreement with the experimental observation from Gutowski \emph{et
al}\cite{2002Gut}. Their HfO$_2$ film was deposited at 823K and then annealed
at 1023K without the formation of any silicides.

It should be emphasized that the thermodynamic stability of HfSi$_2$ in the
HfO$_2$/Si interface depends on the formation energy of HfSiO$_4$ based on the
isothermal sections and the isopleth calculations. The enthalpy of formation
for HfSiO$_4$ is calculated from first-principles calculations since there is
no experimental measurement. To further illustrate this, the reference states
of the enthalpy of formation for HfSiO$_4$ are defined as the two binary metal
oxides (See Eqn. \ref{eqn:hfsio4} and Table \ref{tbl:abinitio}).

\begin{equation}
\label{eqn:hfsio4}%
\Delta H^{HfSiO_4}_{f} = H(HfSiO_4) -\tfrac{1}{2}H(HfO_2) -\tfrac{1}{2}H(SiO_2)
\end{equation}

The current result of the HfSiO$_4$ calculation predicts that HfSi$_2$ is
stable up to 543.53K. However, the uncertainty of the formation enthalpy of
HfSiO$_4$, which originates from the density functional theory itself, is about
$\pm$1 kJ/mol-atom\cite{2002Wol}. Thus, the associated decomposition
temperature of HfSi$_2$ in the HfO$_2$/Si interface varies from 381.92K to
669.76K within the calculated uncertainty of HfSiO$_4$. The recent work from
\citet{2005Miy} found the formation of nanometer-scale HfSi$_2$ dots on the
newly opened void surface produced by the decomposition of HfO$_2$/SiO$_2$
films at the oxide/void boundary in vacuum. However, their result cannot be
directly compared with the current thermodynamic calculations due to the
unknown oxygen partial pressure.

All parameters for the Hf-Si-O system are listed in Table \ref{tbl:parameters}.

\begin{table*}[htb]
\centering%
\caption{Thermodynamic parameters of the Hf-Si-O ternary system (in S.I.
units). Gibbs energies for pure elements and gas phases are respectively from
the SGTE pure elements database\cite{1991Din} and the SSUB
database\cite{1999SGTE}.}
\label{tbl:parameters}%
\begin{tabular}{lll}
\hline %
Phase & Sublattice model & Evaluated description \\
\hline%
Ionic  & $\rm (Hf^{+4})_p(O^{-2},Va)_q$ & %
 $\rm {}^0G^{Ionic~liquid}_{Hf^{+4}:O^{-2}}=2G^{monoclinic}_{HfO_2} +252000-86.798T$ \\
liquid &   & $\rm {}^0G^{Ionic~liquid}_{Hf^{+4}:Va}={}^0G^{Liquid}_{Hf}$ \\
       &   & $\rm {}^0L^{Ionic~liquid}_{Hf^{+4}:O^{-2},Va}=50821+4.203T$ \\
       &   & $\rm {}^1L^{Ionic~liquid}_{Hf^{+4}:O^{-2},Va}=420485-133.300T$ \\
       &   & $\rm {}^2L^{Ionic~liquid}_{Hf^{+4}:O^{-2},Va}=30537$ \\
hcp    & $\rm (Hf)_1(O,Va)_{0.5}$ & $\rm {}^0G^{hcp}_{Hf:Va}={}^0G^{hcp}_{Hf}$ \\
 & & $\rm {}^0G^{hcp}_{Hf:O}={}^0G^{hcp}_{Hf}+0.5{}^0G^{Gas}_{O}-271214+41.560T$\\
       &                          & $\rm {}^0L^{hcp}_{Hf:O,Va}=-31345$ \\
       &                          & $\rm {}^1L^{hcp}_{Hf:O,Va}=-6272$ \\
bcc    & $\rm (Hf)_1(O,Va)_3$     & $\rm {}^0G^{bcc}_{Hf:Va}={}^0G^{bcc}_{Hf}$ \\
 & & $\rm {}^0G^{bcc}_{Hf:O}={}^0G^{hcp}_{Hf}+3{}^0G^{Gas}_{O}-737857+268.540T$\\
       &                          & $\rm {}^0L^{hcp}_{Hf:O,Va}=-981440+20.349T$ \\
HfSiO$_4$ & $\rm (Hf)_1(Si)_1(O)_4$  & $\rm {}^0G^{HfSiO_4}_{Hf:Si:O}= %
G^{monoclinic}_{HfO_2}+G^{quartz}_{SiO_2}-10615+1.313T$\\
\hline %
\end{tabular}
\end{table*}

\section{Summary and conclusion}

The complete thermodynamic description of the Hf-Si-O ternary system is
developed via the hybrid approach of first-principles calculations and CALPHAD
modeling in the present work. In the Hf-O system, special quasirandom
structures have been generated to calculate the enthalpies of mixing of oxygen
and vacancies in the $\alpha$ and $\beta$ solid solutions. Calculated
enthalpies of mixing of $\alpha$-Hf are almost identical to the
model-calculated value whereas those of $\beta$-Hf show significant
discrepancy. In the $\beta$ phase, first-principles calculations could not
retain its original symmetry as bcc due to the strong interaction between the
atoms in the structure. The calculated enthalpies of mixing from SQS's results
are combined with the enthalpies of formations of those hypothetical compounds
calculated from the electronic structure calculations to derive the Gibbs
energy of solid solutions in the Hf-O system.

In the total energy calculation of oxygen gas, vibrational, rotational and
translational degrees of freedom are considered. With the adjusted total energy
of oxygen molecule, the enthalpies of formation from first-principles
calculations for both ordered and disordered phases showed good agreement with
evaluated values.

The Hf-O system has been combined with the Hf-Si and the Si-O systems to
calculate the Hf-Si-O ternary system with the ternary compound HfSiO$_4$
introduced from the first-principles calculations. From the Hf-Si-O
thermodynamic database, phase stabilities pertinent to thin film processing
such as HfO$_2$-SiO$_2$ pseudo-binary, isothermal sections, and isopleth have
been calculated. The thermodynamic calculation results show that the HfO$_2$/Si
interface is stable above 543.53K, which agrees with previous experimental
results. However, due to the uncertainty of HfSiO$_4$ formation energy from
first-principles, the stability of HfSi$_2$ in the HfO$_2$/Si interface is
still in question and further experimental investigation is required.

It can be concluded that the thermodynamic properties of solid phases can be
obtained from first-principles calculations not only for the ordered structures
but also for the solution phases as long as one can find appropriate
geometrical input for phases of interest. Special quasirandom structures for a
substitutional solution phase is one example. However, one should notice that
such a supercell only mimics short-ranged interaction as in metallic alloy
systems. As shown in the present work, SQS's can successfully describe the
mixing behavior between oxygen and vacancies in the $\alpha$ solid solution
where oxygen concentrations are relatively low, but for the oxygen-rich $\beta$
phase such interactions between the electrons at the longer distance become
important and lead to the collapse of its original structure as bcc when the
structure has been fully relaxed.

\section*{Acknowledgements}

This work is funded by the National Science Foundation (NSF) through grants
DMR-0205232/0510180.  First-principles calculations were carried out on the
LION clusters at the Pennsylvania State University supported in part by the NSF
grants (DMR-9983532, DMR-0122638, and DMR-0205232) and in part by the Materials
Simulation Center and the Graduate Education and Research Services at the
Pennsylvania State University.

\bibliographystyle{unsrtnat}

\begin{thebibliography}{44}
\providecommand{\natexlab}[1]{#1} \providecommand{\url}[1]{\texttt{#1}}
\expandafter\ifx\csname urlstyle\endcsname\relax
  \providecommand{\doi}[1]{doi: #1}\else
  \providecommand{\doi}{doi: \begingroup \urlstyle{rm}\Url}\fi

\bibitem[Komarek et~al.(1981)Komarek, Spencer, and Agency.]{1981Kom} K.~L.
    Komarek, P.~J. Spencer, and International Atomic~Energy Agency.
\newblock \emph{Hafnium : physico-chemical properties of its compounds and
  alloys}.
\newblock Atomic Energy Review Special issue ; no 8. International Atomic
  Energy Agency, Vienna, 1981.

\bibitem[Sayan et~al.(2003)Sayan, Garfunkel, Nishimura, Schulte, Gustafsson,
  and Wilk]{2003Say}
S.~Sayan, E.~Garfunkel, T.~Nishimura, W.~H. Schulte, T.~Gustafsson, and G.~D.
  Wilk.
\newblock \emph{J. Appl. Phys.}, 94\penalty0 (2):\penalty0 928--934, 2003.

\bibitem[Hubbard and Schlom(1996)]{1996Hub} K.~J. Hubbard and D.~G. Schlom.
\newblock \emph{J. Mater. Res.}, 11\penalty0 (11):\penalty0 2757--2776, 1996.

\bibitem[Ramanathan et~al.(2003)Ramanathan, McIntyre, Luning, Lysaght, Yang,
  Chen, and Stemmer]{2003Ram}
Shriram Ramanathan, Paul~C. McIntyre, Jan Luning, Patrick~S. Lysaght, Yan Yang,
  Zhiqiang Chen, and Susanne Stemmer.
\newblock \emph{J. Electrochem. Soc.}, 150\penalty0 (10):\penalty0 F173--F177,
  2003.

\bibitem[Gutowski et~al.(2002)Gutowski, Jaffe, Liu, Stoker, Hegde, Rai, and
  Tobin]{2002Gut}
Maciej Gutowski, John~E. Jaffe, Chun-Li Liu, Matt Stoker, Rama~I. Hegde,
  Raghaw~S. Rai, and Philip~J. Tobin.
\newblock \emph{Appl. Phys. Lett.}, 80\penalty0 (11):\penalty0 1897--1899,
  2002.

\bibitem[Waldner and Eriksson(1999)]{1999Wal} Peter Waldner and Gunnar
    Eriksson.
\newblock \emph{CALPHAD}, 23\penalty0 (2):\penalty0 189--218, 1999.

\bibitem[Wang et~al.(2005)Wang, Zinkevich, and Aldinger]{2005Wan} Chong Wang,
    Matvei Zinkevich, and Fritz Aldinger.
\newblock \emph{CALPHAD}, 28\penalty0 (3):\penalty0 281--292, 2005.

\bibitem[Rudy and Stecher(1963)]{1963Rud} E.~Rudy and P.~Stecher.
\newblock \emph{Journal of the Less-Common Metals}, 5\penalty0 (No.
  1):\penalty0 78--89, 1963.

\bibitem[Domagala and Ruh(1965)]{1965Dom} R.~F. Domagala and Robert Ruh.
\newblock \emph{Am. Soc. Metals, Trans. Quart.}, 58\penalty0 (2):\penalty0
  164--75, 1965.

\bibitem[Ruda et~al.(1976)Ruda, Vavilova, Kornilov, Fykin, and
  Panteleev]{1976Rud}
G.~I. Ruda, V.~V. Vavilova, I.~I. Kornilov, L.~E. Fykin, and L.~D. Panteleev.
\newblock \emph{Izvestiya Akademii Nauk SSSR, Neorganicheskie Materialy},
  12\penalty0 (3):\penalty0 461--5, 1976.

\bibitem[Ruh and Patel(1973)]{1973Ruh} Robert Ruh and Vinod~A. Patel.
\newblock \emph{J. Am. Ceram. Soc.}, 56\penalty0 (11):\penalty0 606--7, 1973.

\bibitem[Geller and Corenzwit(1953)]{1953Gel} S.~Geller and E.~Corenzwit.
\newblock \emph{Anal. Chem.}, 25:\penalty0 1774, 1953.

\bibitem[Curtis et~al.(1954)Curtis, Doney, and Johnson]{1954Cur} C.~E. Curtis,
    L.~M. Doney, and J.~R. Johnson.
\newblock \emph{J. Am. Ceram. Soc.}, 37:\penalty0 458--65, 1954.

\bibitem[Adam and Rogers(1959)]{1959Ada} J.~Adam and M.~D. Rogers.
\newblock \emph{Acta Cryst.}, 12:\penalty0 951, 1959.

\bibitem[Boganov et~al.(1965)Boganov, Rudenko, and Makarov]{1965Bog} A.~G.
    Boganov, V.~S. Rudenko, and L.~P. Makarov.
\newblock \emph{Doklady Akademii Nauk SSSR}, 160\penalty0 (5):\penalty0
  1065--8, 1965.

\bibitem[Stacy and Wilder(1975)]{1975Sta} D.~W. Stacy and D.~R. Wilder.
\newblock \emph{J. Am. Ceram. Soc.}, 58\penalty0 (7-8):\penalty0 285--8, 1975.

\bibitem[Wang et~al.(1992)Wang, Li, and Stevens]{1992Wan} J.~Wang, H.~P. Li,
    and R.~Stevens.
\newblock \emph{J. Mater. Sci.}, 27\penalty0 (20):\penalty0 5397--430, 1992.

\bibitem[Massalski(1990)]{1990Mas} T.~B. Massalski.
\newblock \emph{Binary alloy phase diagrams}.
\newblock Binary alloy phase diagrams. ASM International, Materials Park, Ohio,
  2nd edition, 1990.

\bibitem[Speer and Cooper(1982)]{1982Spe} J.~Alexander Speer and Brian~J.
    Cooper.
\newblock \emph{Am. Mineral.}, 67\penalty0 (7-8):\penalty0 804--8, 1982.

\bibitem[Parfenenkov et~al.(1969)Parfenenkov, Grebenshchikov, and
  Toropov]{1969Par}
V.~N. Parfenenkov, R.~G. Grebenshchikov, and N.~A. Toropov.
\newblock \emph{Doklady Akademii Nauk SSSR}, 185\penalty0 (4):\penalty0 840--2,
  1969.

\bibitem[Hirabayashi et~al.(1973)Hirabayashi, Yamaguchi, and Arai]{1973Hir}
    Makoto Hirabayashi, Sadae Yamaguchi, and Tetsuji Arai.
\newblock \emph{J. Phys. Soc. Jpn.}, 35\penalty0 (2):\penalty0 473--81, 1973.

\bibitem[Boureau and Gerdanian(1984)]{1984Bou} G.~Boureau and P.~Gerdanian.
\newblock \emph{J. Phys. Chem. Solids}, 45\penalty0 (2):\penalty0 141--5, 1984.

\bibitem[Komarek and Silver(1963)]{1963Kom} K.~L. Komarek and M.~Silver.
\newblock \emph{Thermodynamics of Nuclear Materials, Proceedings of the
  Symposium on Thermodynamics of Nuclear Materials}, 1962:\penalty0 749--73,
  1963.

\bibitem[Silver et~al.(1963)Silver, Farrar, and Komarek]{1963Sil} M.~D. Silver,
    P.~A. Farrar, and Kurt~L. Komarek.
\newblock \emph{Transactions of the American Institute of Mining, Metallurgical
  and Petroleum Engineers}, 227\penalty0 (4):\penalty0 876--84, 1963.

\bibitem[Kresse and Furthmuller(1996)]{1996Kre} G.~Kresse and J.~Furthmuller.
\newblock \emph{Comput. Mater. Sci.}, 6\penalty0 (1):\penalty0 15--50, 1996.

\bibitem[Kresse and Joubert(1999)]{1999Kre} G.~Kresse and D.~Joubert.
\newblock \emph{Phys. Rev. B}, 59\penalty0 (3):\penalty0 1758--1775, 1999.

\bibitem[Perdew et~al.(1992)Perdew, Chevary, Vosko, Jackson, Pederson, Singh,
  and Fiolhais]{1992Per}
John~P. Perdew, J.~A. Chevary, S.~H. Vosko, Koblar~A. Jackson, Mark~R.
  Pederson, D.~J. Singh, and Carlos Fiolhais.
\newblock \emph{Phys. Rev. B}, 46\penalty0 (11):\penalty0 6671--87, 1992.

\bibitem[McQuarrie(2000)]{McQ00} Donald~A. McQuarrie.
\newblock \emph{Statistical mechanics}.
\newblock Statistical mechanics. University Science Books, Sausalito, Calif.,
  2000.

\bibitem[Zunger et~al.(1990)Zunger, Wei, Ferreira, and Bernard]{1990Zun} Alex
    Zunger, S.~H. Wei, L.~G. Ferreira, and James~E. Bernard.
\newblock \emph{Phys. Rev. Lett.}, 65\penalty0 (3):\penalty0 353--6, 1990.

\bibitem[Jiang et~al.(2004)Jiang, Wolverton, Sofo, Chen, and Liu]{2004Jia} Chao
    Jiang, C.~Wolverton, Jorge Sofo, Long-Qing Chen, and Zi-Kui Liu.
\newblock \emph{Phys. Rev. B}, 69\penalty0 (21):\penalty0 214202/1--214202/10,
  2004.

\bibitem[Shin et~al.(2006)Shin, Arr\'{o}yave, Liu, and van~de Walle]{2006Shi}
    D.~Shin, R.~Arr\'{o}yave, Z.-K. Liu, and A.~van~de Walle.
\newblock \emph{Phys. Rev. B}, 74\penalty0 (2):\penalty0 024204/1--024204/13,
  2006.

\bibitem[Tsuji(1997)]{1997Tsu} Toshihide Tsuji.
\newblock \emph{J. Nucl. Mater.}, 247:\penalty0 63--71, 1997.

\bibitem[Arr\'{o}yave et~al.(2002)Arr\'{o}yave, Kaufman, and Eagar]{2002Arr}
    Raymundo Arr\'{o}yave, Larry Kaufman, and Thomas~W. Eagar.
\newblock \emph{CALPHAD}, 26\penalty0 (1):\penalty0 95--118, 2002.

\bibitem[Redlich and Kister(1948)]{1948Red} O.~Redlich and A.T. Kister.
\newblock \emph{Ind. Eng. Chem.}, 40\penalty0 (2):\penalty0 345--348, 1948.

\bibitem[Hillert et~al.(1985)Hillert, Jansson, Sundman, and Aagren]{1985Hil}
    Mats Hillert, Bo~Jansson, Bo~Sundman, and John Aagren.
\newblock \emph{Metall. Trans. A}, 16A\penalty0 (2):\penalty0 261--6, 1985.

\bibitem[(SGTE)(1999)]{1999SGTE} Scientific Group Thermodata~Europe (SGTE).
\newblock \emph{Thermodynamic Properties of Inorganic Materials}, volume~19 of
  \emph{Landolt-Boernstein New Series, Group IV}.
\newblock Springer, Verlag Berlin Heidelberg, 1999.

\bibitem[Hallstedt(1992)]{1992Hal} Bengt Hallstedt.
\newblock \emph{CALPHAD}, 16\penalty0 (1):\penalty0 53--61, 1992.

\bibitem[Zhao et~al.(2000)Zhao, Bewlay, Jackson, and Chen]{2000Zha} J.~C. Zhao,
    B.~P. Bewlay, M.~R. Jackson, and Q.~Chen.
\newblock \emph{J. Phase Equilib.}, 21\penalty0 (1):\penalty0 40--45, 2000.

\bibitem[Andersson et~al.(2002)Andersson, Helander, Hoglund, Shi, and
  Sundman]{2002And}
J.~O. Andersson, Thomas Helander, Lars Hoglund, Pingfang Shi, and Bo~Sundman.
\newblock \emph{CALPHAD}, 26\penalty0 (2):\penalty0 273--312, 2002.

\bibitem[Chang et~al.(2004)Chang, Shanmugasundaram, Lee, Roman, Wu, Wang,
  Shallenberger, Mumbauer, Grant, Ridley, Dolny, and Ruzyllo]{2004Cha}
K.~Chang, K.~Shanmugasundaram, D.~O. Lee, P.~Roman, C.~T. Wu, J.~Wang,
  J.~Shallenberger, P.~Mumbauer, R.~Grant, R.~Ridley, G.~Dolny, and J.~Ruzyllo.
\newblock \emph{Microelectron. Eng.}, 72\penalty0 (1-4):\penalty0 130--135,
  2004.

\bibitem[Wolverton et~al.(2002)Wolverton, Yan, Vijayaraghavan, and
  Ozolins]{2002Wol}
C.~Wolverton, X.~Y. Yan, R.~Vijayaraghavan, and V.~Ozolins.
\newblock \emph{Acta Mater.}, 50\penalty0 (9):\penalty0 2187--2197, 2002.

\bibitem[Miyata et~al.(2005)Miyata, Morita, Horikawa, Nabatame, Ichikawa, and
  Toriumi]{2005Miy}
Noriyuki Miyata, Yukinori Morita, Tsuyoshi Horikawa, Toshihide Nabatame,
  Masakazu Ichikawa, and Akira Toriumi.
\newblock \emph{Phys. Rev. B}, 71\penalty0 (23):\penalty0 233302/1--233302/4,
  2005.

\bibitem[Dagerhamn(1961)]{1961Dag} Tore Dagerhamn.
\newblock \emph{Acta Chem. Scand.}, 15:\penalty0 214--15, 1961.

\bibitem[Dinsdale(1991)]{1991Din} A.~T. Dinsdale.
\newblock \emph{CALPHAD}, 15\penalty0 (4):\penalty0 317--425, 1991.

\end{thebibliography}

\end{document}